\documentclass[vanilla,keywords,pdf]{iucr}              
\usepackage{amsmath}
\usepackage{graphicx}
\usepackage{amssymb}
\usepackage{layouts}

\newcommand{\customfootnotetext}[2]{{
  \renewcommand{\thefootnote}{#1}
  \footnotetext[0]{#2}}}

\journalcode{J} 
\begin{document}  

\title{Ptychographic X-ray Speckle Tracking with Multi Layer Laue Lens Systems}
\shorttitle{Speckle-tracking}

\cauthor[a]{Andrew J.}{Morgan\textsuperscript{$\dagger$}\customfootnotetext{\dagger}{Current affiliation: ARC Centre of Excellence in Advanced Molecular Imaging, School of Physics, University of Melbourne, Parkville, Victoria 3010, Australia.}}{morganaj@unimelb.edu.au}{}
\author[b]{Kevin T.}{Murray}
\author[b]{Mauro}{Prasciolu}
\author[a]{Holger}{Fleckenstein}
\author[a]{Oleksandr}{Yefanov}
\author[a]{Pablo}{Villanueva-Perez\textsuperscript{$\ddagger$}\customfootnotetext{\ddagger}{Current affiliation: Synchrotron Radiation Research, Lund University, Box 118, 221 00, Lund, Sweden.}}
\author[a]{Valerio}{Mariani}
\author[a]{Martin}{Domaracky}
\author[b]{Manuela}{Kuhn}
\author[a]{Steve}{Aplin\textsuperscript{$\mathsection$}\customfootnotetext{\mathsection}{Current affiliation: European XFEL, Holzkoppel 4, 22869 Schenefeld, Germany.}}
\author[b]{Istvan}{Mohacsi\textsuperscript{$\mathsection$}}
\author[c]{Marc}{Messerschmidt\textsuperscript{$\mathparagraph$}\customfootnotetext{\mathparagraph}{Current affiliation: School of Molecular Sciences, Arizona State University, Tempe, AZ 85287, USA}}
\author[b]{Karolina}{Stachnik}
\author[a]{Yang}{Du\textsuperscript{$\|$}\customfootnotetext{\|}{Current affiliation: School of Physical Science and Technology, ShanghaiTech University, Shanghai, China}}
\author[b]{Anja}{Burkhart}
\author[b]{Alke}{Meents}
\author[d]{Evgeny}{Nazaretski}
\author[d]{Hanfei}{Yan}
\author[d]{Xiaojing}{Huang}
\author[d]{Yong S.}{Chu}
\author[a,e,f]{Henry N.}{Chapman}
\author[b]{Sa\v{s}a}{Bajt}

\aff[a]{CFEL, Deutsches Elektronen-Synchrotron DESY, Notkestraße 85, 22607 Hamburg, Germany}
\aff[b]{DESY, Notkestrasse 85, 22607 Hamburg, Germany}
\aff[c]{National Science Foundation BioXFEL Science and Technology Center, 700 Ellicott Street, Buffalo, NY 14203, USA}
\aff[d]{National Synchrotron Light Source II, Brookhaven National Laboratory, Upton, NY 11973, USA}
\aff[e]{Centre for Ultrafast Imaging, Luruper Chaussee 149, 22761 Hamburg, Germany}
\aff[f]{Department of Physics, University of Hamburg, Luruper Chaussee 149, 22761 Hamburg, Germany}


\shortauthor{Andrew J. Morgan \textit{et al.}}


\maketitle 

\begin{synopsis}
Simultaneous wavefront metrology and sample projection imaging with multi layer Laue lenses using the ptychographic x-ray speckle tracking technique. We present results from three experiments.
\end{synopsis}


\begin{abstract}
The ever-increasing brightness of synchrotron radiation sources demands
improved x-ray optics to utilise their capability for imaging and probing
biological cells, nano-devices, and functional matter on the nanometre scale
with chemical sensitivity. Hard x-rays are ideal for high-resolution imaging
and spectroscopic applications due to their short wavelength, high penetrating
power, and chemical sensitivity. The penetrating power that makes x-rays useful
for imaging also makes focusing them technologically challenging. 
Recent developments in layer deposition techniques that have enabled
the fabrication of a series of highly focusing x-ray lenses, known as wedged
multi layer Laue lenses. Improvements to the lens design and fabrication
technique demands an accurate, robust, in-situ and at-wavelength
characterisation method. To this end, we have developed a modified form of
the speckle-tracking wavefront metrology method, the ptychographic x-ray speckle tracking method, which is capable of operating with highly divergent wavefields. A useful by-product of this method, is that it also provides high-resolution and aberration-free projection images of extended specimens. We report on three separate experiments using this method, where we have resolved ray path angles to within $4\;$nano-radians with an imaging resolution of $45\;$nm (full-period).
This method does not require a high degree of
coherence, making it suitable for lab based x-ray sources. Likewise it is
robust to errors in the registered sample positions making it suitable for
x-ray free-electron laser facilities, where beam pointing fluctuations can be
problematic for wavefront metrology.
\end{abstract}



\section{Introduction}



\begin{figure}
\includegraphics[width=8.88cm]{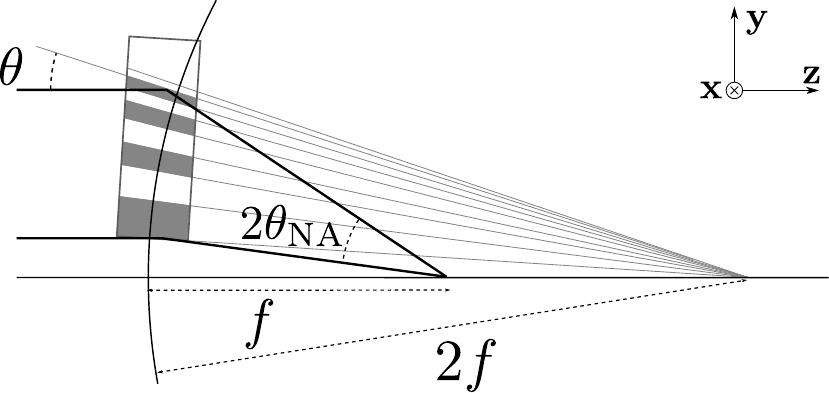}
\caption{A wedged multilayer Laue lens of focal length f is constructed from layers whose spacing follows the zone-plate condition. To achieve high efficiency the lens must be thick, in which case diffraction is a volume effect described by dynamical diffraction. In this case the layers should be tilted to locally obey Bragg’s law, which places them normal to a circle of radius 2f. 
} 
\label{fig:MLL}
\end{figure}

In 2015 \citeasnoun{Morgan2015} reported on the use of a lens for one dimensional focusing of hard x-rays, with a photon energy of $22\;$keV. This lens was made by alternately depositing two materials with layer periods that follow the Fresnel zone-plate condition and then slicing the structure approximately perpendicular to the layers to the desired optical thickness. By varying the tilt of the layers throughout the stack, so that the Bragg and zone plate conditions are simultaneously fulfilled for every layer, large focusing angles can be achieved with uniform efficiency. Such a structure is referred to as a wedged Multi-layer Laue Lens (MLL) \cite{Yan2014}, which is fabricated by the use of a masked magnetron sputtering technique, and is schematically illustrated in Fig. \ref{fig:MLL}. 

Errors in the wavefront produced by the wedged MLL were characterised using a pseudo one dimensional ptychographic algorithm. This analysis revealed a defect in the lens that produced two distinct regions along the layer stack, each with a different focal length. Further studies revealed that the defect was caused by a transition in the material pair from amorphous to crystalline phase for layer periods of about $5.5\;$nm \cite{Bajt2018}. By switching to a new materiel pair (tungsten carbide and silicon carbide) the phase transition could be avoided, allowing for a larger lens stack with greater focusing power. This illustrates the importance of wavefront metrology as a diagnostic tool for the iterative development of new optical elements. 
\onecolumn
\begin{figure}
\includegraphics[width=\textwidth]{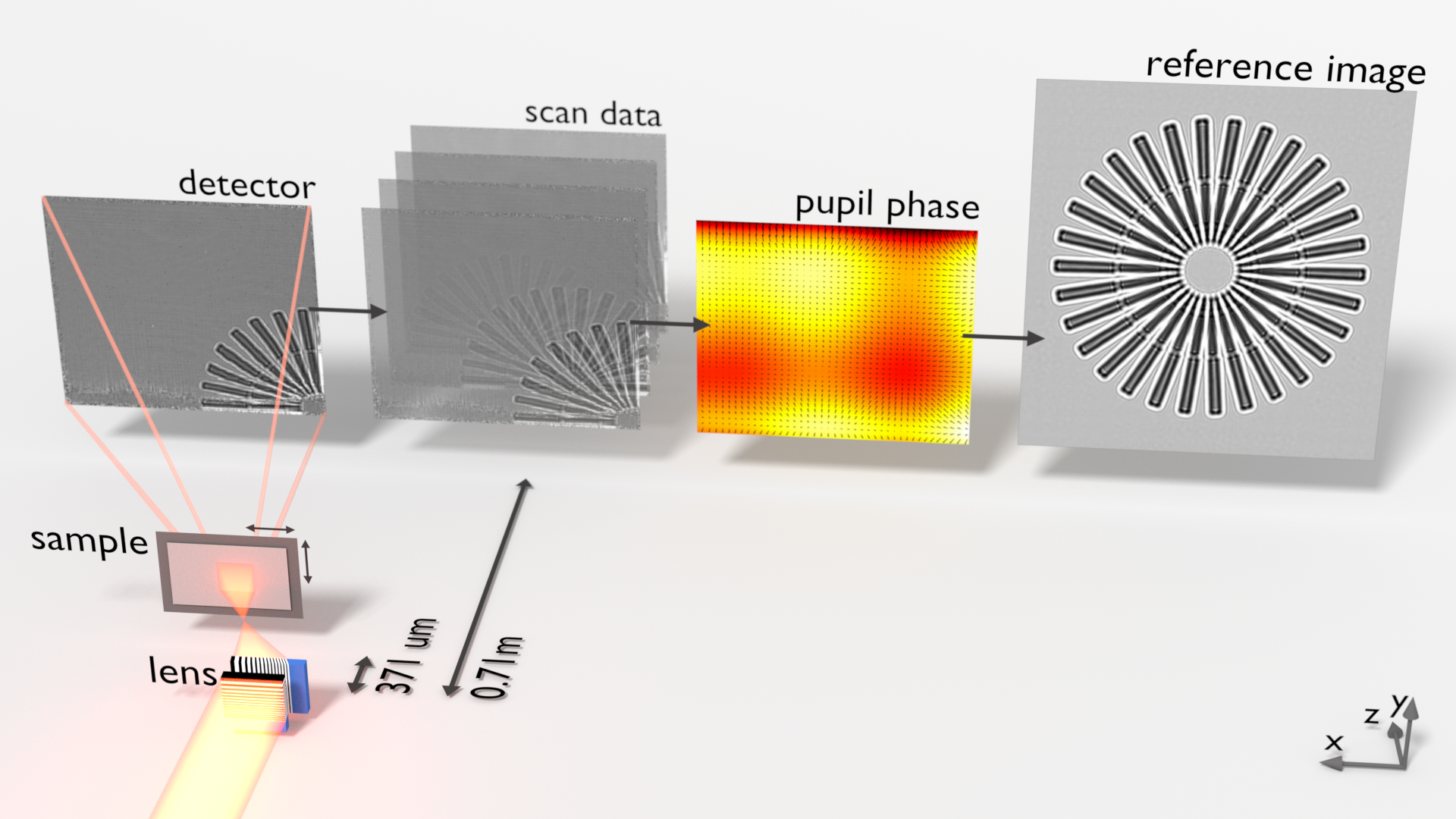}
\caption{Illustration of the Ptychographic-XST method. The beamline 
    illumination was focused (off-axis) in 2D by two crossed and wedged MLLs. 
    The Siemens star sample was placed $371\;\mu$m downstream of the 
    focal plane. Images were recorded on a CCD pixel array detector $0.71\;$m 
    downstream of the focus. The scan data consists of 49 shadow images, 
    recorded as the sample was translated across the beam profile. The phase 
    and reference image maps were refined iteratively. } 
\label{fig:overview}
\end{figure}
\twocolumn
\noindent 
Ptychography is a powerful tool for wavefront metrology, as it allows for the simultaneous recovery of the complex-valued wavefront produced by the lens and the complex-valued transmission function of the sample which is scanned across the wavefront, typically near the focal plane of the lens, with diffraction limited resolution \cite{Chapman1996, Rodenburg2007, Thibault2009}.  
The high resolution is a result of the fact that ptychography often employs a fully coherent model\footnote{Methods for dealing with partial coherence, that can characterised by a few dominant modes, have been successfully developed for ptychography \cite{Thibault2013, Pelz2014}. Nevertheless, each mode is treated in a full coherent way, consistent with the original (single-mode) ptychographic approach.} for the wavefront propagation from the sample to the detector plane, with few approximations beyond paraxial illumination, a thin specimen and a high degree coherence for the imaging system.

However, ptychography can present difficulties in its implementation, in part because the coherent model of the imaging system can be sensitive to errors in the estimated model parameters. It can also be computationally demanding to perform the required number of iterative steps in the reconstruction algorithm, which can be exacerbated by the large number of diffractions patterns in some ptychographic datasets. Furthermore, determination of the root cause of a failed reconstruction, for example, bad detector readings, sample stage jitter, x-ray source incoherence or algorithm parameters, can be difficult due to the complicated relationship between the measured diffraction intensities and the recovered wavefronts. For example, although the wavefront reconstruction in \citeasnoun{Morgan2015} took only a few hours to complete, this calculation was preceded by many months of work identifying detector artefacts, exploring reconstruction parameters and checking the uniqueness of the output. 

Since the work of \cite{Berujon2012a} and \cite{Morgan2012a}, X-ray Speckle-tracking Techniques (XSTs) have emerged as a viable tool for wavefront metrology applications. This method is based on near-field speckle based imaging, where the 2D phase gradient of
a wavefield can be recovered by tracking the displacement of localised 
``speckles'' between an image and a reference image produced in the projection hologram of an object with a random phase/absorption profile. Additionally, 
XST can be employed to measure the phase profile of an object's transmission 
function. Thanks to the simple experimental set-up, high angular sensitivity, 
and compatibility with low coherence sources, this method has since been 
actively developed for use in synchrotron and laboratory light sources; see 
\citeasnoun{Zdora2018} for a recent review. 

As part of an ongoing project to develop and improve the fabrication and performance of wedged MLLs for imaging \cite{Prasciolu2015a, Murray2019a}, we have developed a modified form of the XST suitable for highly divergent illumination conditions \cite{Morgan2019}. This method, the Ptychographic X-ray Speckle-tracking Technique (PXST), adopts an experimental geometry and iterative update algorithm similar to that employed in many ptychographic applications. Under ideal imaging conditions, the PXST method will not achieve the same (diffraction limited) sample imaging resolution or phase sensitivity that could be achieved via ptychographic approaches. However, we show that it is possible to recover images with large magnified factors, of around 2000 or more, and thus PXST can provide sufficiently high phase sensitivity and imaging resolution for many applications. Based on a pseudo-geometric approximation for the propagation of light from the sample exit-surface to the detector plane, the source of errors in the recovered wavefronts can be localised to individual intensity measurements leading to a more transparent and more easily diagnosed reconstruction process. We present PXST results from three separate experiments, each with a different sample, effective magnification and defocus distance.


\section{Wavefront analysis}

The experiment set-up and processing pipeline are roughly equivalent for each experiment, as illustrated in Fig. \ref{fig:overview}. In this configuration the sample was placed a distance $z_1$ downstream of the 2D beam focus, which was formed using two crossed and wedged MLLs (one MLL to focus vertically and the other horizontally). The focal length of the lens closest to the focus was reduced by its distance from the other lens so that the focal points for the two MLLs meet in the same plane. A total of $N$ images ($I_n$) where then recorded on a detector placed a distance $z$ downstream of the sample, as the sample was translated in a 2D grid pattern a distance $\Delta \textbf{x}_n$ in the x-y plane (perpendicular to the optical axis for the n'th image). If $z_1$ is sufficiently large, then the images formed on the detector resemble shadow images of the sample, which are variously called Gabor or in-line holograms, near-field images, phase contrast images \textit{etc.} depending on the specific application and properties of the sample (for the rest of this article we shall refer to such images simply as shadow images). 

In the ideal case, for a thin sample, a lens system without any aberrations, ignoring diffraction from a lens aperture and for large $z_1$, the lens will produce a spherical wavefront and it can be shown that the observed shadow image will be equivalent to a defocused and magnified image of the sample ($I_\text{ref}$), such that $I_n(\textbf{x}, z) = M^{-2}I_\text{ref}(\textbf{x}/M - \Delta \textbf{x}_n, \bar{z})$, where the magnification factor $M$ is given by $(z_1+z)/z_1$ and the effective defocus $\bar{z}$ is given by $z z_1 / (z_1 + z)$. In \cite{Morgan2019}, this principle was generalised to incorporate the divergent illumination formed by a non-ideal lens system, so that:
\begin{align}
 I_n(\textbf{x}) \approx W(\textbf{x}) I_\text{ref}(\textbf{u}(\textbf{x}) - \Delta \textbf{x}_n, \bar{z}), \label{eq:I}
\end{align}
where $W(\textbf{x})$ is the ``whitefield image'', the intensity distribution measured on the detector without the presence of the sample, and $\textbf{u}(\textbf{x})$ is a 2D vector field that captures both the average magnification of the image (due to the global phase curvature of the illumination) and the geometric distortions (arising from the finite aperture and lens aberrations) in the shadow image, given by:
\begin{align}\label{eq:u}
 \textbf{u}(\textbf{x}) = \textbf{x} - \frac{\lambda z}{2\pi} \nabla \Phi(\textbf{x}),
\end{align}
where $\lambda$ is the wavelength of the radiation, $\nabla = (\partial/\partial x, \partial / \partial y)$ is the transverse gradient operator and $\Phi$ is the phase of the wavefield produced by the lens system in the detector plane (in the absence of the sample).


\onecolumn
\begin{table}
\centering
    \caption{}
\begin{tabular}{lrrr}
    \textbf{Sample} & \textbf{Siemens Star} & \textbf{Diatom} & \textbf{Diatom-subregion} \\ 
    beamline                     & NSLS-II (HXN)         & PETRA III (P11) & PETRA III (P11) \\
    energy (keV)                 & $16.7$                & $16.3$ & $16.3$ \\
    focus-detector distance (m)  & $0.71$                & $1.32$ & $1.32$ \\
    focus-sample distance (mm)   & $0.371$               & $2.22$ & $0.55$ \\
    detector                     & Merlin                & Lambda & Lambda \\
    detector grid (ROI)          & $407\times 365$       & $359\times 401$ & $359\times 401$ \\
    physical pixel area ($\mu$m$^2$)  & $55\times 55$    & $55\times 55$   & $55\times 55$ \\
    effective pixel area (nm$^2$)     & $30\times 28$    & $93\times 92$   & $24\times 24$ \\
    average magnification        & $1917$                & $595$           & 2308 \\
    effective defocus (mm)       & $0.37$                & $2.21$          & 0.57 \\
    sample scan grid             & $20\times 20$         & $11\times 11$   & $11\times 11$ \\
    sample scan step size ($\mu$m) & $0.63$              & $10.00$         & $0.20$ \\
    exposure time (s)            & $1$                   & $5$             & 0.005  \\
    iterations                   & 3                     & 3               & 3 \\
    angular resolution (nrad)    & 6.3                   & 20.0            & 3.4 
\end{tabular}

\label{table:parameters}
\end{table}
\twocolumn
Using Eqs. \ref{eq:I} and \ref{eq:u} and the set of shadow images ($I_n$), the wavefront formed by the two MLLs in the detector plane, given by the phase ($\Phi$) and intensity ($W$), as well as the un-distorted, magnified and defocused image of the sample, which we call the ``reference image'' $I_\text{ref}$, were recovered by tracking the local displacement of features formed in each of the shadow images according to the recipe described in \cite{Morgan2019} using a speckle-tracking software package\footnote{\label{note1}Found here: www.github.com/andyofmelbourne/speckle-tracking}. In this method, initial estimates for $\nabla \Phi$, $I_\text{ref}$ and $\Delta \textbf{x}$ are iteratively refined until the sum squared error between the measurements and the forward model (given by the forward model in Eq. \ref{eq:I}) is minimised. The experiment data for each of the results shown here are available on the CXIDB. The analysis presented here can be replicated by following the tutorial sections on the software website$^{2}$.
The parameters for each experiment are summarised in table \ref{table:parameters}.

\subsection{Image reconstruction with the example of the Siemens star sample}

\begin{figure}
\includegraphics[width=8.88cm]{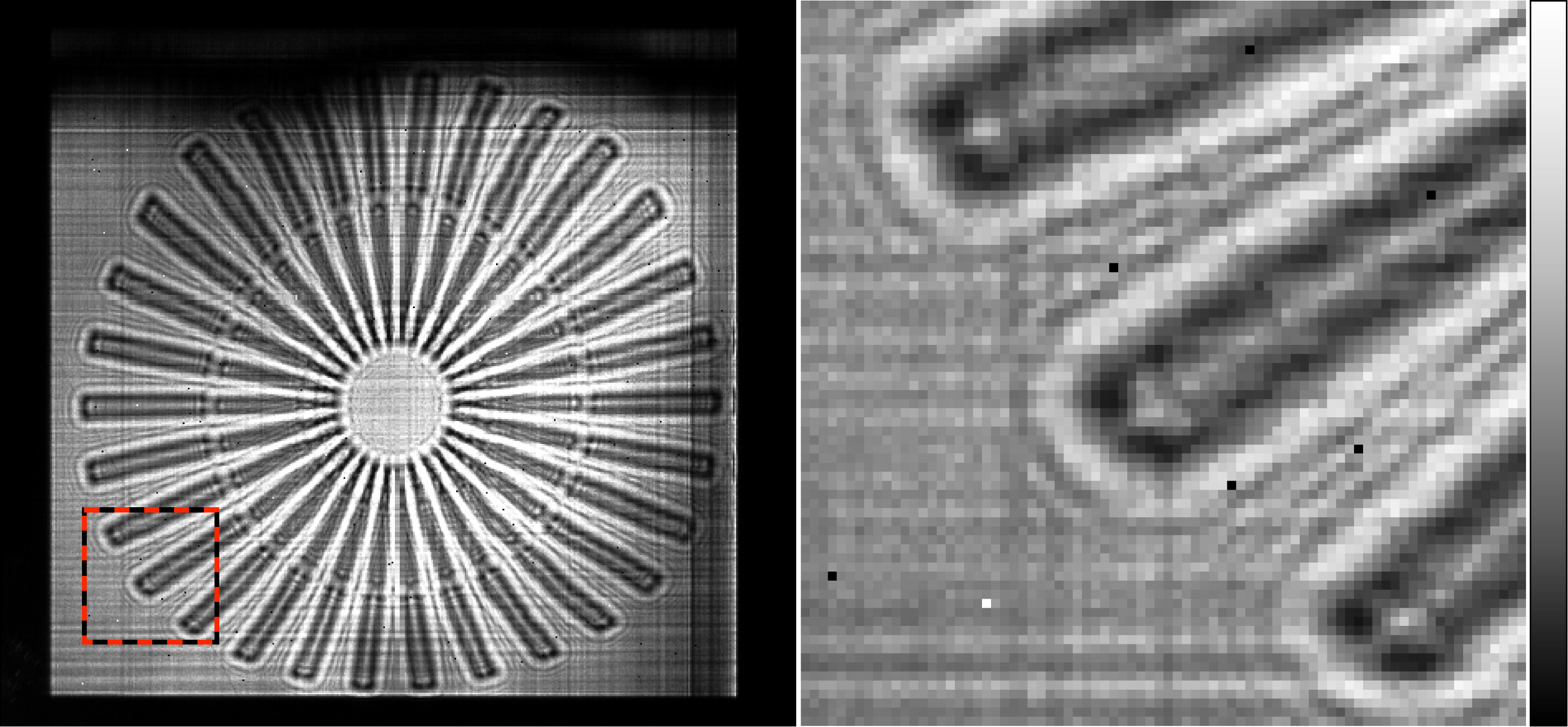}
\caption{\textbf{Left:} Raw detector image of the Siemens star shadow (480x438 pixels). The un-focused beam was blocked by a beam stop placed beyond the bottom left of this figure. The checkered red and black box outlines the region shown on the right at a higher image magnification. The linear colour scale is displayed at the very right and ranges from 0 (black) to 3000 (white) photon counts.  } 
\label{fig:frame}
\end{figure}

\noindent For this experiment, shadow images of a Siemens star test sample were recorded at the NSLS-II HXN beamline\cite{Nazaretski2014, Nazaretski2017}. Figure \ref{fig:frame} (left) shows one of the 400 shadow images recorded as part of the scan. To achieve a 2D focus we would ideally use two MLLs, one to focus vertically and the other horizontally, that are optimised for the same photon energy. In this experiment however, we had one lens that was optimal at $16.7\;$keV and the other at $16.9\;$keV. We decided to operate at $16.9\;$keV. Because of this mismatch of $0.2\;$keV, the vertically focusing MLL does not focus x-rays with uniform efficiency across the entire physical aperture. 
This results in the tapered fall-off in diffraction intensity near the top of the figure corresponding to higher diffraction angles, the optical axis is located beyond the bottom left of the figure. The horizontally focusing MLL provided efficiency is nearly uniform across the entire pupil region along the horizontal direction. In addition to scattering from the sample and the faint cross-hatch pattern (which we speculate are due to small local variations in the layer periods), there are also intensity variations across the image caused by the non-uniform illumination incident on the MLL lens system from up-stream optical elements. 
The Siemens star test sample with a total diameter of $10\;\mu$m and consists of 30 radial ``spokes'' with circular cuts at two radial positions. It is constructed from gold with a projected thickness in the range $0.5$ to $1\;\mu$m with a minimum feature size of $50\;$nm. The geometry of the Siemens star helps to visualise the effect of the low order aberrations in the lens system on the observed shadow images. These aberrations led to low spatial frequency geometric distortions that break the approximate circular symmetry of the image, which is evident in Fig. \ref{fig:frame} (left). 
\begin{figure}
\includegraphics[width=8.88cm]{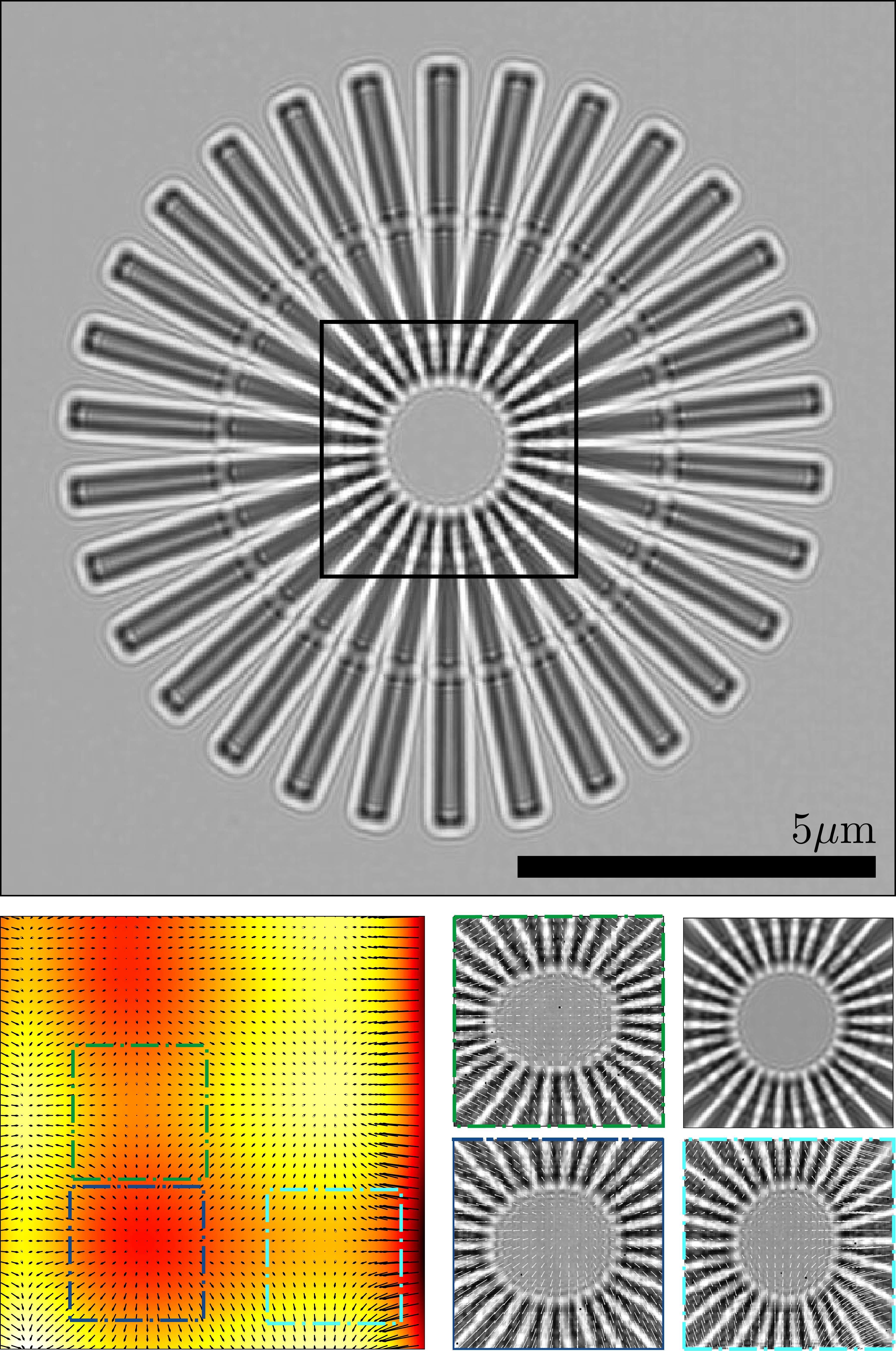}
\caption{\textbf{Top:} Reference image ($I_\text{ref}$) of the Siemens star test sample. \textbf{Bottom left:} Phase profile of the pupil function $\Phi$ (colour scale), overlayed with a quiver plot of the retrieved phase gradient $\nabla \Phi$ vector field (scaled to pixel units). \textbf{Bottom right:} Four views of the central region of the Siemens star: At the top right is the undistorted view (as outlined in black in the top panel), the remaining three panels show this feature as it appears in different locations on the detector array (after division by $W(\textbf{x})$) corresponding to the regions indicated by like-coloured outlines in the left panel. } 
\label{fig:distortions}
\end{figure}
\noindent To the right we show a magnified view of the region of interest. Here, we can observe approximately three Fresnel fringes generated by the sharp outer edges of the Siemens star spokes. This is the same fringe structure one would observe by illuminating the sample with plane wave illumination and recording an image on a detector placed a distance $\bar{z} = 0.37\;$m downstream of the object and magnified by a factor of $M=1917$. The effective Fresnel number is then given by $\bar{\text{F}}=X^2/(\lambda \bar{z})$, where $X$ is the full period spatial frequency of a feature in the sample. In the present case we have $\bar{F}=0.09$, corresponding to the smallest feature size of $X=50\;$nm. 
\begin{figure}
\includegraphics[width=8.88cm]{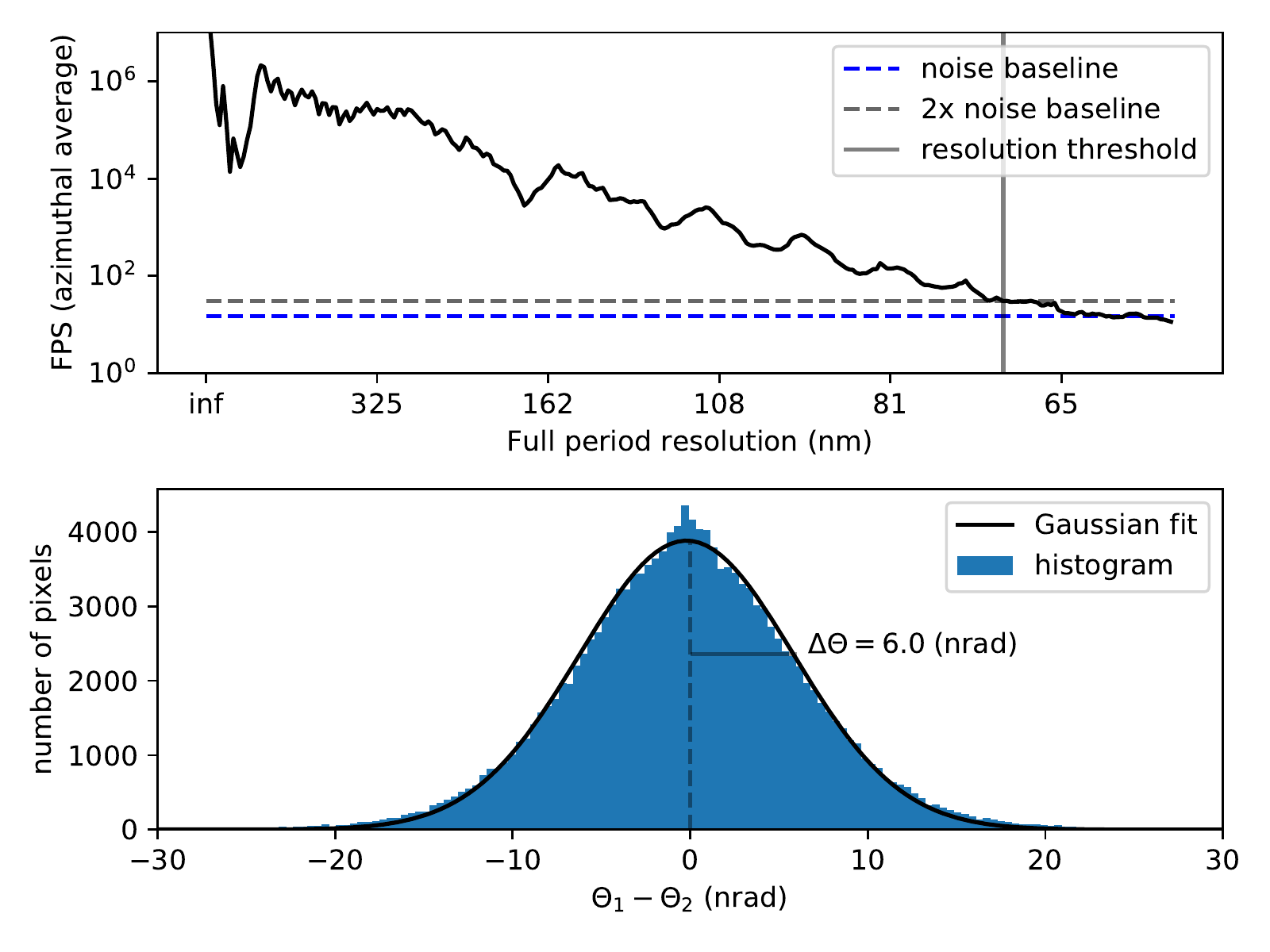}
    \caption{\textbf{Top}: The azimuthal average of the Fourier Power Spectrum (FPS) of the recovered reference image of the Siemens star sample. The FPS is obtained by taking the mod square of the Fourier transform of $I_\text{ref}$. The blue dashed line shows the noise floor which was estimated by taking the average of the FPS over the last 30 values. The resolution cut off (grey vertical line) is given by the resolution at which the FPS is equal to twice the noise floor (black dashed line). \textbf{Bottom}: Histogram of the difference between the recovered wavefront angles (detector plane) from each of the split-1/2 datasets (blue bar chart). The solid black line shows the Gaussian model fit with a standard deviation of $0.6\;$nrad.} 
\label{fig:res}
\end{figure}
The whitefield image ($W$), was set to the median value at each pixel on the detector over the 400 measurements. A more direct approach would have been to record an image after completely removing the sample from the incident wavefield. We found however, that the former strategy led to superior results. We speculate that this is due to low frequency drifts in either the positioning or the upstream illumination of the MLL system, leading to small variations in the intensity profile of the beam. Naturally, these drifts also occur during the acquisition time of the dataset and could limit the viability of this method in cases where the duration of the experiment far exceeds the duration of stability for the imaging system. 

The initial estimate for the gradient of the wavefield in the detector plane ($\nabla \Phi$) was set to:
\begin{align}\label{eq:z1xy}
 \nabla \Phi(\textbf{x}) &= \frac{2\pi}{\lambda}(\frac{x}{z_1^x}, \frac{y}{z_1^y}),
\end{align}
where $z_1^x$ and $z_1^y$ are the distances between the sample plane and the horizontal and vertical focal planes of the lens system respectively. Note that for an astigmatic lens system, $z_1^x \neq z_1^y$. Estimates for $z_1^x$ and $z_1^y$ were obtained, in turn, by fitting a set of parameters in a forward model for the power spectrum of the data, obtained by summing the mod-square of the Fourier transform of each image. The Fresnel fringes present in each image produce a nearly circular ring pattern in the cumulative power spectrum, known as ``Thon rings'', where the shape and spacing of the rings provide estimates for defocus and astigmatism. This algorithm$^{2}$
, was adapted from the program CTFFIND4 \cite{Rohou2015} (developed for use on cryo-electron microscopy micrographs).

\begin{figure}
\includegraphics[width=8.88cm]{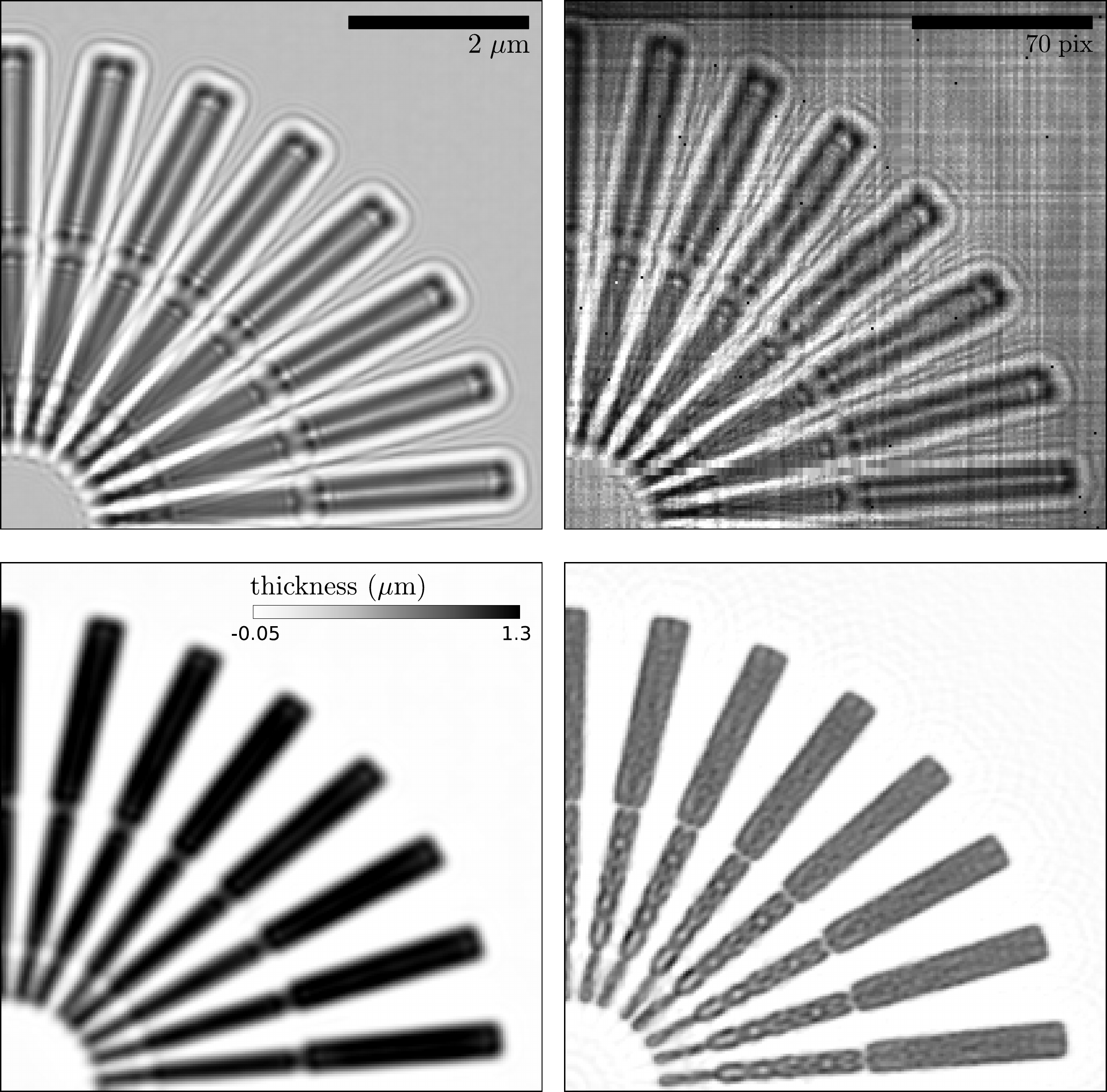}
    \caption{\textbf{Top left}: Subregion of the reference image reconstruction. The linear colour scale ranges from 0 (black) to 1.6 (white). \textbf{Top right}: Subregion of image 250 in the dataset, without any preprocessing. The linear colour scale ranges from 0 (black) to 4000 (white) photon counts. \textbf{Bottom left}: TIE reconstruction of the same subregion as in the top left panel. \textbf{Bottom right}: CTF inverted reconstruction of the same region. The colour scale in the same as in the bottom left panel. } 
\label{fig:tie}
\end{figure}

In the top panel of Fig. \ref{fig:distortions} we show the reconstructed reference image of the sample ($I_\text{ref}$). We note that this is not a real-space image of the sample but rather a magnified view of the defocused image. Correctly reconstructed, this reference image will be free of the geometric aberrations present in each of the measured images, and indeed, this appears to be the case here. The direct (real-space) imaging resolution is limited by the effective defocus distance, so that point-like features will produce overlapping spots at a separation distance less than $331\;$nm (Rayleigh criterion), rather than the de-magnified pixel size of $28\;$nm. This is the separation distance between the inner edges of the spokes of the Siemens star when the first minimum of the edge's Fresnel fringes overlaps with the brighter zero'th order maximum of the adjacent edge. Another measure of resolution, is the Fourier Power Spectrum (FPS) cut-off frequency, which is given by the highest spatial frequencies in an image above the signal to noise level. The FPS is graphed in Fig. \ref{fig:res} (top panel), with the vertical black line indicating the full period resolution of the image at $70\;$nm, or a half period resolution of $35\;$nm, approximately $20\%$ greater than the de-magnified pixel size. 

In Fig. \ref{fig:tie} we show two real-space reconstructions of the Siemens star's projected mass (bottom left and right panels). For a sample constructed from a single material, with a constant density, and a linear approximation to Beer's law, the projected mass is equal to the thickness, or height of the sample above the substrate.
Both were recovered from the reference image (top left panel) and can be compared with a raw diffraction image, shown in the top right panel. In the bottom row, we display the thickness profile recovered via the Transport of Intensity Equation (TIE) and via Contrast Transfer Function (CTF) inversion, in the left and right panels respectively, using the X-TRACT software package \cite{Gureyev2011}.
We note that both methods are not ideal in the present case: the TIE algorithm works best for large Fresnel numbers and the CTF inversion is ideal for weak phase objects. Nevertheless, the ends of the ``spokes'' near the centre of the Siemens star, with a separation distance $\approx 158\;$nm can clearly be distinguished in both images, which is an improvement on the direct (real-space) resolution of the reference image. 
\begin{figure}
\includegraphics[width=8.88cm]{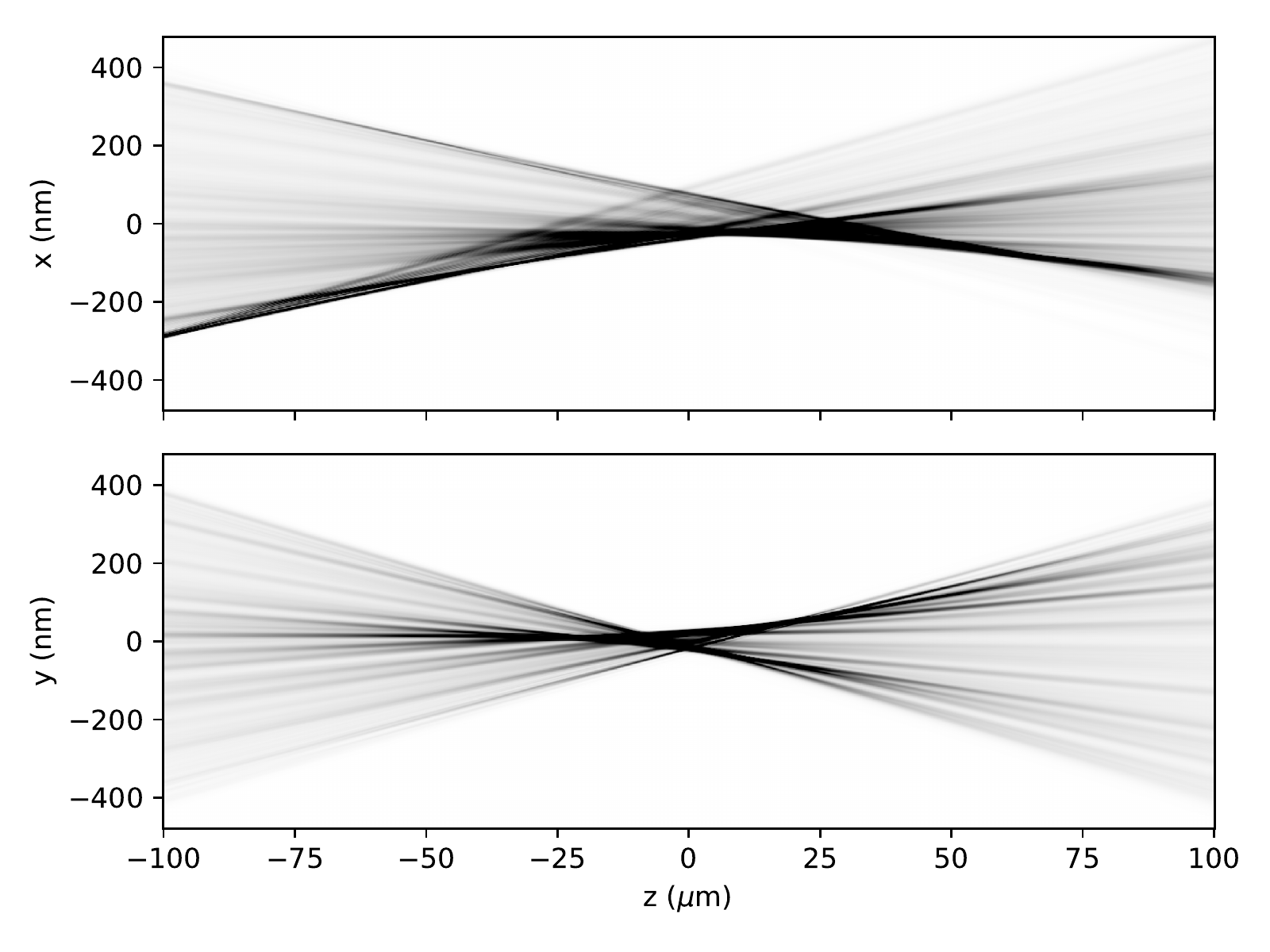}
    \caption{Projection of the wavefront intensity profile near the focal plane of the lens system, along the $x$-axis (\textbf{top}) and the $y$-axis (\textbf{bottom}). The linear colour scale ranges from 0 (white) to 1 (black) in arbitrary units.} 
\label{fig:probe}
\end{figure}
The vector field $\textbf{u}(\textbf{x}) - \Delta \textbf{x}_n$ defines the mapping between each point in the $n$'th image ($I_n(\textbf{x})$) to a point in the reference image, see Eq. \ref{eq:I}. The phase gradients were obtained from $\textbf{u}$ via Eq. \ref{eq:u}, using the formalism in \cite{Morgan2019}, and are shown in the bottom left panel of Fig. \ref{fig:distortions}. Here we display the phase gradients, after removing the global shift and magnification factors, as a black quiver plot, scaled to pixel units. 
In order to further illustrate the effect of the geometric aberrations, beyond the overall magnification, caused by the phase gradients in the lens phase profile, we display a magnified view of the central region of the Siemens star (see the black box in the top panel of Fig. \ref{fig:distortions}) as it appears in three different shadow images. The region where this feature appears for each of the three images are illustrated by the coloured square outlines shown in the bottom left panel. In the bottom right panel we show the corresponding regions for each of these images. To increase the contrast, we have divided the images by the whitefield image ($I_n / W$). In the top right sub-panel, we also show the same region of the recovered reference image. Here one can clearly observe local variations in the degree of magnification, along the $x$ and / or $y$ axis, depending on the position of the sample within the incident wavefield.

In the bottom left panel we show the residual phase profile of the MLL lens system (colour map). The residual phase profile is obtained after removing the constant, linear and quadratic components of the global phase profile, which correspond to an overall phase constant, a tilt term and the defocus aberrations respectively. By removing these terms, it is possible to perceive the small deviations in the phase from an (ideal) quadratic profile. Armed with this phase profile, we could then numerically propagate the wavefield to the region near the focal plane of the lens, as shown in Fig. \ref{fig:probe}. These results were obtained after 3 iterations of the PXST update algorithm. For each iteration, we refined the initial estimates for the sample stage translations. The ``irrotational constraint'' on the phase gradients was also enforced (see section 5 of \cite{Morgan2019}). 

In appendix \ref{sec:comparison} we show a comparison of the recovered wavefront phase from a separate PXST experiment and a ptychographic experiment taken with the sample placed nearer to the focal plane. Both results show qualitative agreement, however the root-mean-squared difference is much greater than we would predict if the ptychographic result is considered to be the ground truth.

The local angular distribution of the wavefront rays, in the plane of the detector, are given by $\Theta = \frac{\lambda}{2\pi}\nabla \Phi$. In the ideal case, the smallest resolvable angular deviation of a ray (the angular sensitivity) is given by $\Delta \Theta = \delta_\text{pix}\sigma_\text{det} /M$, where $\sigma_\text{det}$ is the width of the point spread function of the detector (greater than or equal to the physical pixel size) and $\delta_\text{pix}$ is the fractional reduction in the effective pixel size due to numerical interpolation. In the present case, setting $\sigma_\text{det}\approx 55\mu$m and $\delta_\text{pix}<1$, we have $\Delta \Theta < 29\;$nrad. 

In order to estimate the achieved angular resolution, we randomly assigned each pixel of each image to one of two datasets. Keeping the reconstructed reference map and sample stage positions from the original reconstruction, we then repeated the reconstruction of the phase gradients independently for each of the two datasets. This process is only possible due to the high degree of redundancy in the original data. A histogram of the difference between the two reconstructions, shown in the bottom panel of Fig. \ref{fig:res}, provides an estimate for the underlying uncertainty in the recovered $\Theta$ values. The standard deviation of the difference $\Theta_1 - \Theta_2$, yields $\Delta \Theta \approx 6.0\;$nrad, which suggests a $\delta_\text{pix}$-value of less than $2/10$. 
This shows that from the redundancy of data, caused by measurements at any given location in the wave with many positions of the object, that one is able to interpolate angular deviations to a small fraction of a pixel. 
The angular distribution is related to the phase profile via 2D integration $\Phi(\textbf{x}) = \frac{2\pi}{\lambda} \iint \Theta(\textbf{x}) d\textbf{x}$ and propagating the uncertainties yields an estimate for the phase sensitivity of $\Delta \Phi \approx 0.065\;$rad ($0.01$ waves).


\onecolumn
\begin{figure}
\includegraphics[width=\textwidth]{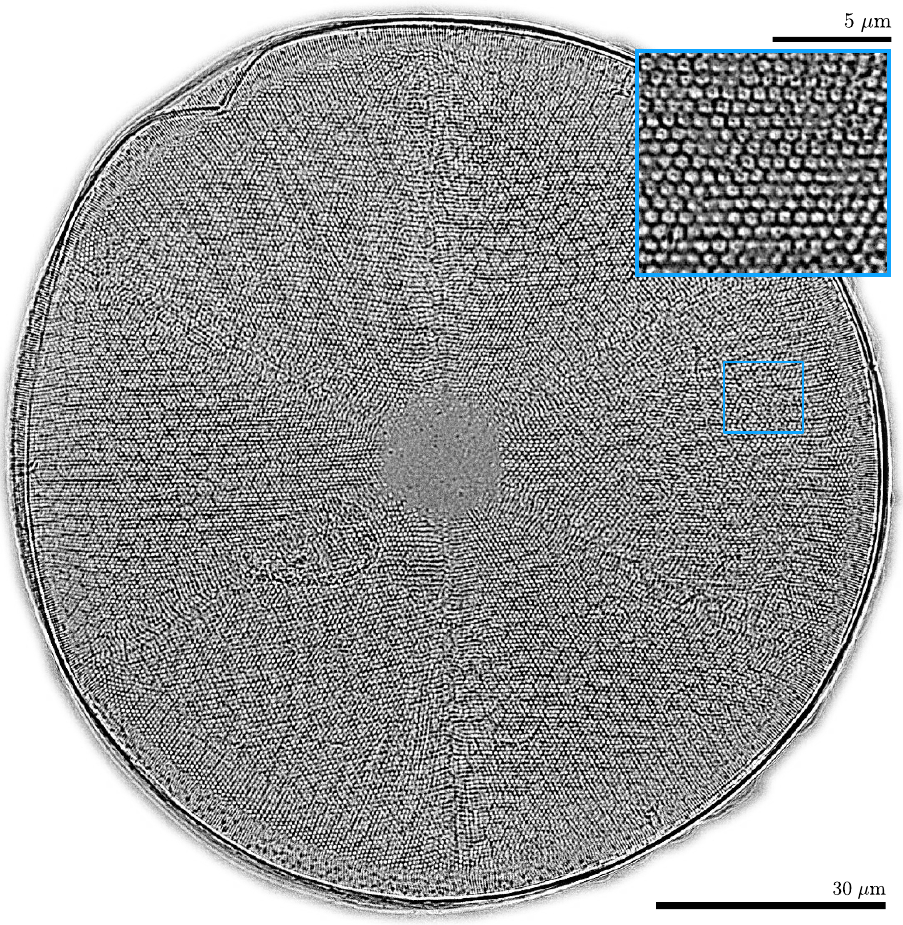}
    \caption{Sample image map of a diatom. The linear grey scale colour map ranges from 0.92 (black) to 1.08 (white). The reconstructed area outside of the Diatom's region of interest has been masked. The magnified pixel area is $93\times 93\;$nm$^2$. The field of view of the image is $122\times 120\;\mu$m$^2$ ($1320 \times 1290\;$pixels). Fine details in the sub-structure of the Diatom are visible in this phase contrast projection image, which are otherwise obscured by the surface of the sample in scanning electron micrograph images. \textbf{Top right:} magnified image map of a subregion of the Diatom. The field of view is $95 \times 107\;\mu$m ($414 \times 466\;$pixels). The small blue rectangle indicates the scale of the inset with respect to the larger image of the diatom.} 
\label{fig:diatom}
\end{figure}
\twocolumn

\subsection{Diatom sample}

For this experiment, the biomineralized shell of a marine planktonic diatom was placed on a silicon nitride membrane and scanned across the wavefield $2.22\;$mm down stream of the lens focus. In contrast to the Siemens star experiment, the effective defocus and magnification (see Table \ref{table:parameters}) are such that only first order Fresnel fringes are visible across the majority of the reference image. For this reason we did not use the Thon rings to provide initial estimates for $z_1^x$ and $z_1^y$. Instead, we set $z_1=z_1^x=z_1^y$ in Eq. \ref{eq:z1xy} and chose the value of $z_1$ which minimised the sum squared error after many trials over a range of $z_1$ values. Errors in the initial estimates for $z_1$, $z_1^x$ and $z_1^y$ will lead to additional defocus aberrations in the recovered phase map, which can then be removed as needed. If these errors are too large however, the algorithm may take many more iterations (or fail completely) to converge. 

In contrast to the previous experiment, only a fraction (roughly 1/9'th) of the object is visible in the field of view for each image. The reference image is shown in Fig. \ref{fig:diatom}, obtained after three iterations of the PXST algorithm. 

This diatom was collected from the Antarctic sea and its shell is made from a complex network of nanostructured silica with an exceptional strength to weight ratio, despite being produced under low temperature and pressure conditions. The circular shell of the diatom is constructed from 6 azimuthal segments, which extend in a dome-like fashion out of the page for the orientation shown in Fig. \ref{fig:diatom}. The boundary of these segments can be observed as the 6 radial creases, extending from the edge of the inner circle to the outer rim of the sample. This 6-fold symmetry is a motif that is repeated throughout the diatom structure, see for example, the approximate hexagonal packing of the small ``white dots'' with a diameter of about $5\;\mu$m. In another scan (discussed in the next section) taken with the sample closer to the focus, a more detailed view of these ``white dots'' can be seen. This more magnified view of the diatom is displayed in the top right corner of the figure, and one can see that these dots are themselves hexagonal in shape with what appear to be hollow depressions in the centre. 

The estimated angular sensitivity for this reconstruction is $20\;$nrad, which is approximately $3.2$ times greater than for the Siemens star reconstruction. This result is consistent with the corresponding decrease in the average magnification by a factor of $3.3$, from $1917$ (Siemens star) to $595$ (diatom). The direct (real-space) imaging resolution was $410\;$nm (Rayleigh criterion), while the FPS cut-off frequency was $259\;$nm, with a half period resolution of $130\;$nm which is $40\%$ greater than the de-magnified pixel size.

\subsection{Diatom subregion}

For this experiment the sample was moved closer to the focal plane of the lens, from $2.22\;$mm in the previous section to $0.57\;$mm here. This corresponds to an increase in the magnification by a factor of $3.9$, from 595 to 2308. As discussed in \cite{Morgan2019}, the upper limit to the magnification factor for this particular technique is governed by the smallest distance between the focal plane and the sample such that the diffraction remains in the near-field imaging regime. For larger magnification factors, with the sample closer to the focal plane, the rapidly oscillating phase and intensity of the illuminating wavefield lead to significant errors in the speckle tracking approximation of Eq. \ref{eq:I}. Here however, another difficulty was encountered relating to the pseudo translational symmetry of the diatom structure at this magnification.
%
\begin{figure}
\includegraphics[width=8.88cm]{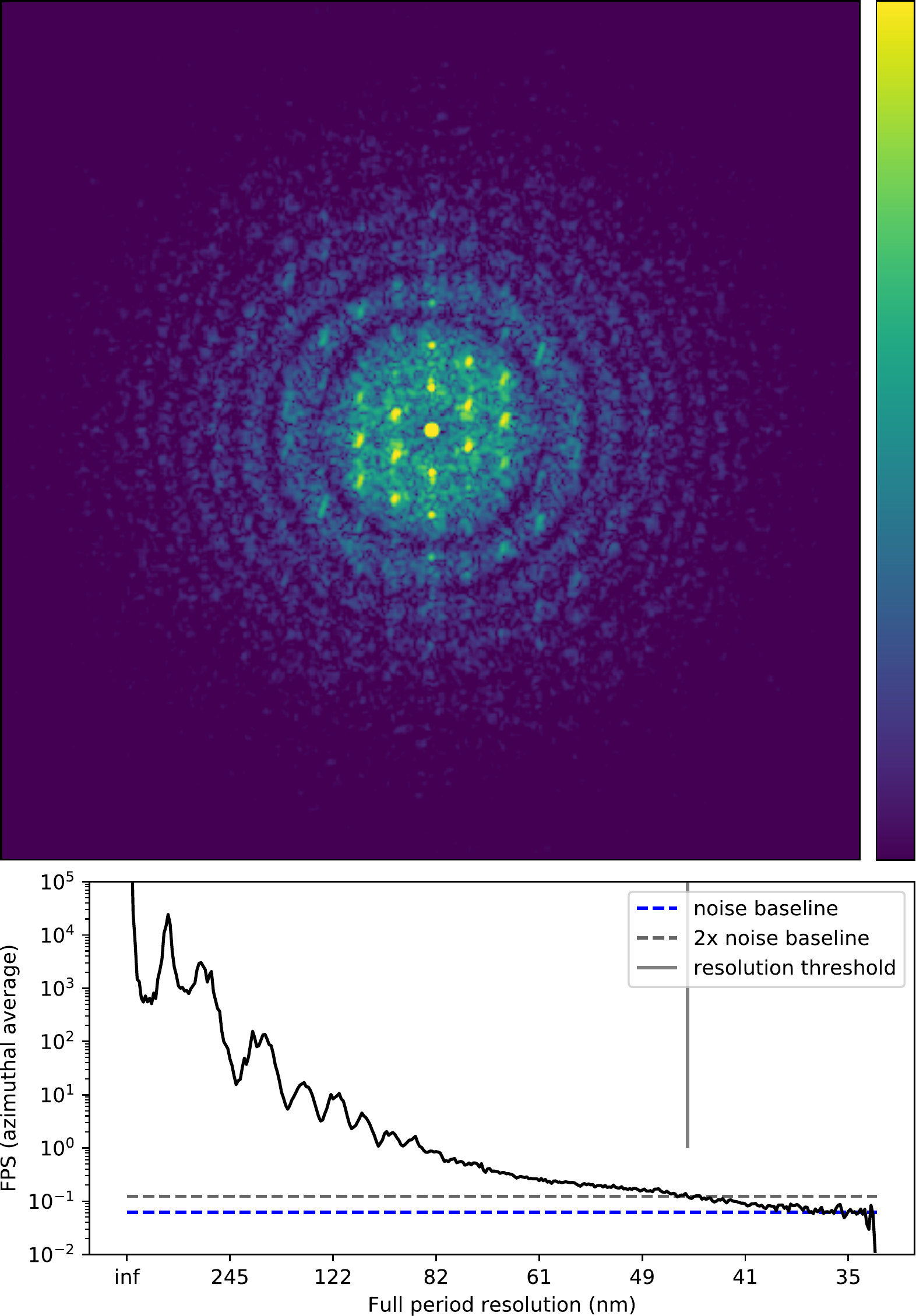}
    \caption{\textbf{Top}: Image of the FPS of the diatom subregion. The full period spatial frequency at the left edge of the image is $48\;$nm$^{-1}$. To avoid artefacts from the sharp edges of the real-space image (as shown in subpanel of Fig. \ref{fig:diatom}) the FPS was filtered with a Gaussian window function with a standard deviation of $2.4\;\mu$m. Before display the FPS was raised to the power 0.1, in order to reveal the Thon rings underneath the much stronger peaks from the hexagonal lattice. \textbf{Bottom}: Azimuthal average of the FPS, with a cut-off frequency corresponding to a full period resolution of $45\;$nm (half period resolution of $22.5\;$nm, $5\%$ smaller than the de-magnified pixel size). } 
\label{fig:FPSd}
\end{figure}
The FPS of the reference image in the top panel of Fig. \ref{fig:FPSd} shows an hexagonal array of points overlaid on top of the much weaker Thon rings, which (again) arise because the reference image is a defocused image of the sample's exit-surface wave. The location of the peaks reveal the reciprocal lattice of the real-space structure, which is approximately hexagonal with a primitive lattice constant of $\approx 601\;$nm. This approximate translational symmetry is undesirable in PXST because of the possibility of miss-registering features between each recorded image and the reference by an amount equal to the lattice constant.

\begin{figure}
\includegraphics[width=8.88cm]{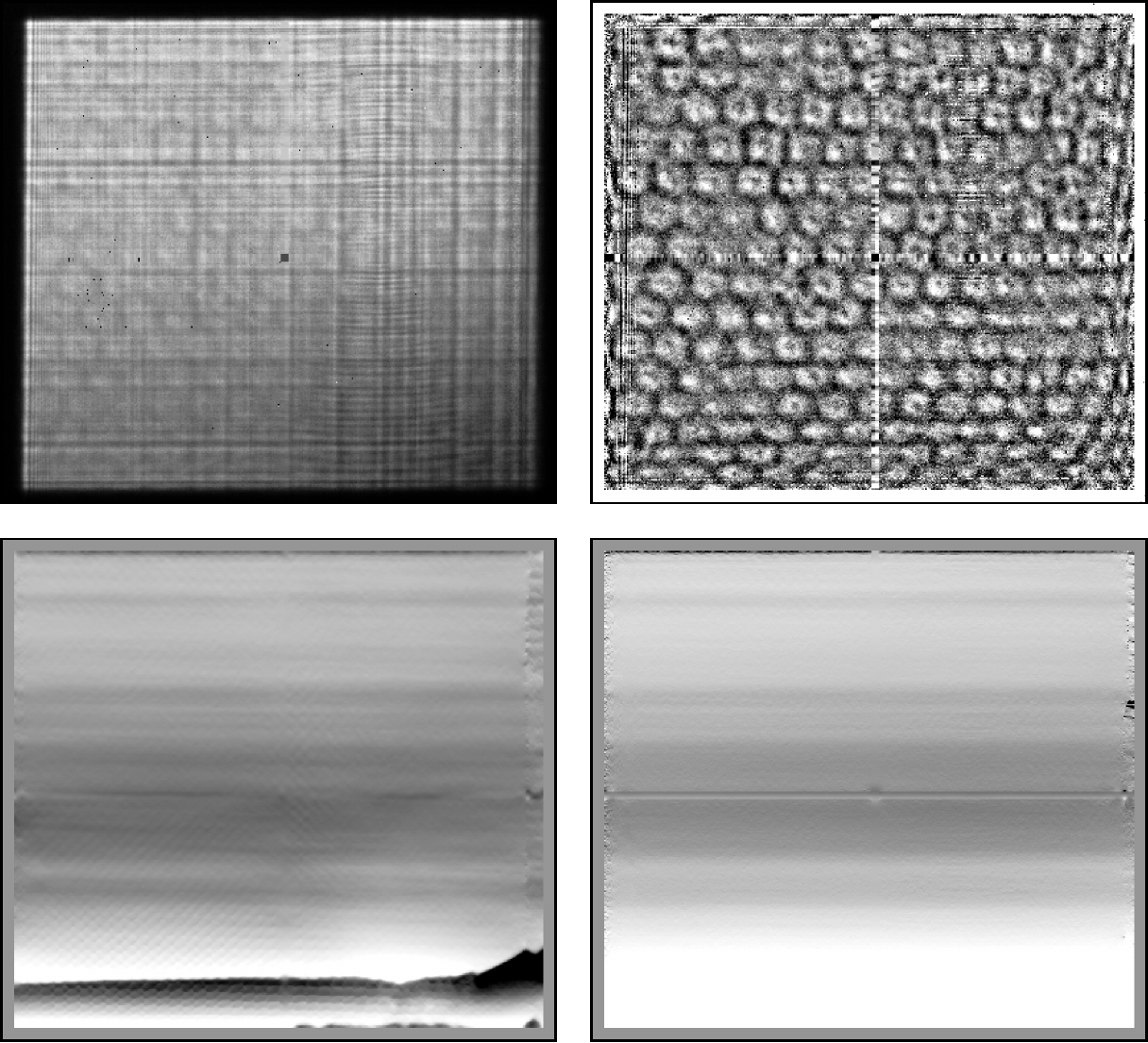}
    \caption{\textbf{Top left}: Image 50 of the 121 recorded shadow images. This image spans diffraction angles of $15\times 17\;$mrads. The linear colour scale ranges from 0 (black) to 2000 (white) photon counts. \textbf{Top right}: The same image divided by the whitefield ($W$), colour scale ranges from 0.9 to 1.2. \textbf{Bottom left}: The recovered pixel mapping between the recorded images and the reference image $\textbf{u}(\textbf{x})$ (in pixel units), colour scale ranges from ranges from -10 (black) to 10 (white) pixel shifts. \textbf{Bottom right}: The recovered pixel mapping when employing regularisation during the reconstruction, same colour scale as bottom left.}
\label{fig:u}
\end{figure}

In the bottom left panel of Fig. \ref{fig:u} we show a failed reconstruction of the pixel mapping between the recorded images (one of which is shown in the top left panel) and the reference. At the bottom of the image one can see a horizontal step-like reduction in the mapping function from white to black, corresponding to reduction of 20 pixels. When scaled to physical units, this drop corresponds exactly to the hexagonal lattice spacing of the diatom substructure. In order to overcome this problem, we chose to regularise the recovered pixel shifts by convolving them with gaussian kernel at each iteration. The standard deviation of this kernel was reduced linearly from 20 pixels to 0 pixels as the iterations progressed. In this way sharp deviations in the mapping function were prevented from forming early in the reconstruction process. The result of this regularisation procedure is shown in the bottom right panel of the figure, where the step like artefact is no longer present in the reconstructed pixel mapping. 





\section{Discussion}

In this article we have demonstrated the use of the PXST on three experiment datasets. In each case, both the illuminating wavefront and a highly magnified, un-distorted phase-contrast image of the sample were recovered. The main benefit of PXST over other speckle tracking techniques, for example, the Unified Modulated Pattern Analysis (UMPA) approach of \citeasnoun{Zdora}, the geometric flow algorithm of \citeasnoun{Paganin2018} and the original XST technique of \citeasnoun{Berujon2012a}, is that it is able to deal with highly divergent illumination. This allows for the comparatively large magnification factors (e.g. $2308$ for the diatom subregion), which leads to a corresponding increase in the achievable phase sensitivity ($3.4\;$nrad) and image resolution ($45\;$nm full-period). Conversely, PXST does not provide a direct (real-space) image of the sample's phase, absorption or the so called ``dark-field'' profiles.

Another approach that is suitable for highly divergent illumination is the X-ray Speckle-Scanning technique of \citeasnoun{Berujon2012a}, which provides a phase sensitivity proportional to the step size of the sample translations. In PXST however, the phase sensitivity does not depend on the step size, making it suitable for a broader range of experiment facilities. 

With the high NA, efficient, hard x-ray optics provided by the wedged MLLs used here, the footprint of the beam on the sample is greater for a fixed magnification factor than would otherwise be the case. This increases the throughput of the imaging method, by a factor proportional to the square of the increase in the NA.

We have also demonstrated that PXST does not require an additional diffuser in the beam path and we expect that a wide variety of samples could be used as a wavefront sensing device -- although a dense random object such as a diffuser should reduce the number of required images.

In future, we hope to develop the PXST algorithm for use in ``cone-beam tomography'', a geometry where the illumination diverges significantly as it passes through the object.

\section{Acknowledgements}

We acknowledge Lars Gumprecht, Julia Maracke, Siegfried Imlau (CFEL), Sabrina Bolmer, Sven Korseck, Janning Meinert, Florian Pithan, David Pennicard, Andre Rothkirch and Heinz Graafsma (DESY) for various contributions. 
Andrzej Andrejczuk (Univ. of Bialystok, Poland) assisted with theoretical aspects of the propagation of light through the MLLs. 
Timur E. Gureyev assisted with the TIE and CTF based reconstructions of the Siemens star. 
Christian Hamm, from the Alfred Wegener Institute, Helmholtz Centre for Polar and Marine Research provided the diatom sample. 
Karolina Stachnik acknowledges the Joachim-Herz Stiftung.
Funding for this project was provided by: the Australian Research Council Centre of Excellence in Advanced Molecular Imaging (AMI), the Gottfried Wilhelm Leibniz Program of the DFG and the NSF award 1231306. This research used the HXN beamline of the National Synchrotron Light Source II, a U.S. Department of Energy (DOE) Office of Science User Facility operated for the DOE Office of Science by Brookhaven National Laboratory under Contract No. DE-SC0012704.

\bibliographystyle{iucr}
\bibliography{speckle_tracking_iucr}

@article{Nazaretski2017,
author = {Nazaretski, E. and Yan, H. and Lauer, K. and Bouet, N. and Huang, X. and Xu, W. and Zhou, J. and Shu, D. and Hwu, Y. and Chu, Y. S.},
doi = {10.1107/S1600577517011183},
issn = {16005775},
journal = {Journal of Synchrotron Radiation},
month = {nov},
number = {6},
pages = {1113--1119},
publisher = {International Union of Crystallography},
title = {{Design and performance of an X-ray scanning microscope at the Hard X-ray Nanoprobe beamline of NSLS-II}},
volume = {24},
year = {2017}
}

@article{Nazaretski2014,
author = {Nazaretski, E. and Huang, X. and Yan, H. and Lauer, K. and Conley, R. and Bouet, N. and Zhou, J. and Xu, W. and Eom, D. and Legnini, D. and Harder, R. and Lin, C. H. and Chen, Y. S. and Hwu, Y. and Chu, Y. S.},
doi = {10.1063/1.4868968},
issn = {10897623},
journal = {Review of Scientific Instruments},
number = {3},
publisher = {American Institute of Physics Inc.},
title = {{Design and performance of a scanning ptychography microscope}},
volume = {85},
year = {2014}
}

@inproceedings{Gureyev2011,
author = {Gureyev, Timur E. and Nesterets, Yakov and Ternovski, Dimitri and Thompson, Darren and Wilkins, Stephen W. and Stevenson, Andrew W. and Sakellariou, Arthur and Taylor, John A.},
booktitle = {Advances in Computational Methods for X-Ray Optics II},
doi = {10.1117/12.893252},
isbn = {9780819487513},
issn = {0277786X},
month = {sep},
pages = {81410B},
publisher = {SPIE},
title = {{Toolbox for advanced x-ray image processing}},
volume = {8141},
year = {2011}
}

@article{Morgan2019,
author = {Morgan, Andrew J. and Quiney, Harry M. and Bajt, Sa{\v{s}}a and Chapman, Henry N.},
journal = {JAC - special issue on ptychography (submitted)},
number = {special issue Ptychography},
title = {{Ptychographic X-ray Speckle Tracking}},
year = {2019}
}

@article{Thibault2013,
author = {Thibault, Pierre and Menzel, Andreas},
doi = {10.1038/nature11806},
issn = {00280836},
journal = {Nature},
month = {feb},
number = {7435},
pages = {68--71},
title = {{Reconstructing state mixtures from diffraction measurements}},
volume = {494},
year = {2013}
}

@article{Pelz2014,
author = {Pelz, Philipp M. and Guizar-Sicairos, Manuel and Thibault, Pierre and Johnson, Ian and Holler, Mirko and Menzel, Andreas},
doi = {10.1063/1.4904943},
issn = {00036951},
journal = {Applied Physics Letters},
month = {dec},
number = {25},
publisher = {American Institute of Physics Inc.},
title = {{On-the-fly scans for X-ray ptychography}},
volume = {105},
year = {2014}
}

@article{Prasciolu2015a,
author = {Prasciolu, M. and Leontowich, A. F. G. and Krzywinski, J. and Andrejczuk, A. and Chapman, H. N. and Bajt, S.},
doi = {10.1364/ome.5.000748},
issn = {2159-3930},
journal = {Optical Materials Express},
month = {apr},
number = {4},
pages = {748},
publisher = {The Optical Society},
title = {{Fabrication of wedged multilayer Laue lenses}},
volume = {5},
year = {2015}
}

@article{Rohou2015,
author = {Rohou, Alexis and Grigorieff, Nikolaus},
doi = {10.1016/J.JSB.2015.08.008},
issn = {1047-8477},
journal = {Journal of Structural Biology},
month = {nov},
number = {2},
pages = {216--221},
publisher = {Academic Press},
title = {{CTFFIND4: Fast and accurate defocus estimation from electron micrographs}},
url = {https://www-sciencedirect-com.ezp.lib.unimelb.edu.au/science/article/pii/S1047847715300460?via{\%}3Dihub},
volume = {192},
year = {2015}
}

@article{Morgan2012a,
author = {Morgan, Kaye S. and Paganin, David M. and Siu, Karen K. W.},
doi = {10.1063/1.3694918},
issn = {0003-6951},
journal = {Applied Physics Letters},
month = {mar},
number = {12},
pages = {124102},
publisher = {American Institute of Physics},
title = {{X-ray phase imaging with a paper analyzer}},
url = {http://aip.scitation.org/doi/10.1063/1.3694918},
volume = {100},
year = {2012}
}

@article{Chapman1996,
author = {Chapman, Henry N.},
doi = {10.1016/S0304-3991(96)00084-8},
issn = {0304-3991},
journal = {Ultramicroscopy},
month = {dec},
number = {3-4},
pages = {153--172},
publisher = {North-Holland},
title = {{Phase-retrieval X-ray microscopy by Wigner-distribution deconvolution}},
url = {https://www-sciencedirect-com.ezp.lib.unimelb.edu.au/science/article/pii/S0304399196000848},
volume = {66},
year = {1996}
}

@article{Berujon2012a,
author = {Berujon, Sebastien and Wang, Hongchang and Sawhney, Kawal},
doi = {10.1103/PhysRevA.86.063813},
issn = {1050-2947},
journal = {Physical Review A},
month = {dec},
number = {6},
pages = {063813},
publisher = {American Physical Society},
title = {{X-ray multimodal imaging using a random-phase object}},
url = {https://link.aps.org/doi/10.1103/PhysRevA.86.063813},
volume = {86},
year = {2012}
}

@article{Zdora2018,
author = {Zdora, Marie-Christine and Zdora and Marie-Christine},
doi = {10.3390/jimaging4050060},
issn = {2313-433X},
journal = {Journal of Imaging},
month = {apr},
number = {5},
pages = {60},
publisher = {Multidisciplinary Digital Publishing Institute},
title = {{State of the Art of X-ray Speckle-Based Phase-Contrast and Dark-Field Imaging}},
url = {http://www.mdpi.com/2313-433X/4/5/60},
volume = {4},
year = {2018}
}

@article{Murray2019a,
author = {Murray, Kevin T. and Pedersen, Anders F. and Mohacsi, Istvan and Detlefs, Carsten and Morgan, Andrew J. and Prasciolu, Mauro and Yildirim, Can and Simons, Hugh and Jakobsen, Anders C. and Chapman, Henry N. and Poulsen, Henning F. and Bajt, Sa{\v{s}}a},
doi = {10.1364/oe.27.007120},
issn = {1094-4087},
journal = {Optics Express},
month = {mar},
number = {5},
pages = {7120},
publisher = {Optical Society of America},
title = {{Multilayer Laue lenses at high X-ray energies: performance and applications}},
url = {https://www.osapublishing.org/abstract.cfm?URI=oe-27-5-7120},
volume = {27},
year = {2019}
}

@article{Paganin2018,
author = {Paganin, David M. and Labriet, H{\'{e}}l{\`{e}}ne and Brun, Emmanuel and Berujon, Sebastien},
doi = {10.1103/PhysRevA.98.053813},
issn = {2469-9926},
journal = {Physical Review A},
month = {nov},
number = {5},
pages = {053813},
publisher = {American Physical Society},
title = {{Single-image geometric-flow x-ray speckle tracking}},
url = {https://link.aps.org/doi/10.1103/PhysRevA.98.053813},
volume = {98},
year = {2018}
}

@article{Bajt2018,
author = {Bajt, Sa{\v{s}}a and Prasciolu, Mauro and Fleckenstein, Holger and Domarack{\'{y}}, Martin and Chapman, Henry N. and Morgan, Andrew J and Yefanov, Oleksandr and Messerschmidt, Marc and Du, Yang and Murray, Kevin T and Mariani, Valerio and Kuhn, Manuela and Aplin, Steven and Pande, Kanupriya and Villanueva-Perez, Pablo and Stachnik, Karolina and Chen, Joe P. J. and Andrejczuk, Andrzej and Meents, Alke and Burkhardt, Anja and Pennicard, David and Huang, Xiaojing and Yan, Hanfei and Nazaretski, Evgeny and Chu, Yong S and Hamm, Christian E},
doi = {10.1038/lsa.2017.162},
issn = {2047-7538},
journal = {Light: Science {\&} Applications},
month = {mar},
number = {3},
pages = {17162},
publisher = {Nature Publishing Group},
title = {{X-ray focusing with efficient high-NA multilayer Laue lenses}},
url = {http://www.nature.com/doifinder/10.1038/lsa.2017.162},
volume = {7},
year = {2018}
}

@techreport{Martin,
author = {Martin, Andrew V. and Berntsen, Peter and Kozlov, Alex and Kewish, Cameron and Nugent, Keith and Greaves, Tamar and Persson, Ingmar},
pages = {0--7},
title = {{A.V. Martin Probing 3D nanostructure of liquids and early-stage crystallization in the lipidic cubic phase}}
}

@article{Zdora,
author = {Zdora, Marie Christine and Thibault, Pierre and Zhou, Tunhe and Koch, Frieder J. and Romell, Jenny and Sala, Simone and Last, Arndt and Rau, Christoph and Zanette, Irene},
doi = {10.1103/PhysRevLett.118.203903},
issn = {10797114},
journal = {Physical Review Letters},
number = {20},
title = {{X-ray Phase-Contrast Imaging and Metrology through Unified Modulated Pattern Analysis}},
url = {https://journals.aps.org/prl/pdf/10.1103/PhysRevLett.118.203903},
volume = {118},
year = {2017}
}

@article{Morgan2015,
author = {Morgan, Andrew J. and Prasciolu, Mauro and Andrejczuk, Andrzej and Krzywinski, Jacek and Meents, Alke and Pennicard, David and Graafsma, Heinz and Barty, Anton and Bean, Richard J. and Barthelmess, Miriam and Oberth{\"{u}}r, Dominik and Yefanov, Oleksandr M. and Aquila, Andrew and Chapman, Henry N. and Bajt, Sa{\v{s}}a},
doi = {10.1038/srep09892},
isbn = {doi:10.1038/srep09892},
issn = {2045-2322},
journal = {Scientific reports},
number = {1},
pages = {9892},
pmid = {26030003},
publisher = {Nature Publishing Group},
title = {{High numerical aperture multilayer Laue lenses.}},
url = {http://www.nature.com/srep/2015/150601/srep09892/full/srep09892.html},
volume = {5},
year = {2015}
}

@article{Thibault2009,
author = {Thibault, Pierre and Dierolf, Martin and Bunk, Oliver and Menzel, Andreas and Pfeiffer, Franz},
issn = {0304-3991},
journal = {Ultramicroscopy},
number = {4},
pages = {338--343},
title = {{Probe retrieval in ptychographic coherent diffractive imaging}},
url = {http://www.sciencedirect.com/science/article/pii/S0304399108003458},
volume = {109},
year = {2009}
}

@article{Yan2014,
author = {Yan, Hanfei and Conley, Ray and Bouet, Nathalie and Chu, Yong S},
issn = {0022-3727},
journal = {Journal of Physics D: Applied Physics},
number = {26},
pages = {263001----},
title = {{Hard x-ray nanofocusing by multilayer Laue lenses}},
url = {http://stacks.iop.org/0022-3727/47/i=26/a=263001},
volume = {47},
year = {2014}
}

@article{Rodenburg2007,
author = {Rodenburg, John M and Hurst, A. C. and Cullis, A. G.},
doi = {10.1016/j.ultramic.2006.07.007},
isbn = {0304-3991},
issn = {03043991},
journal = {Ultramicroscopy},
number = {2-3},
pages = {227--231},
pmid = {16959428},
title = {{Transmission microscopy without lenses for objects of unlimited size}},
volume = {107},
year = {2007}
}


\appendix

\section{Comparison of the recovered wavefront via ptychography and PXST}\label{sec:comparison}

\begin{figure}
\includegraphics[width=0.8\textwidth]{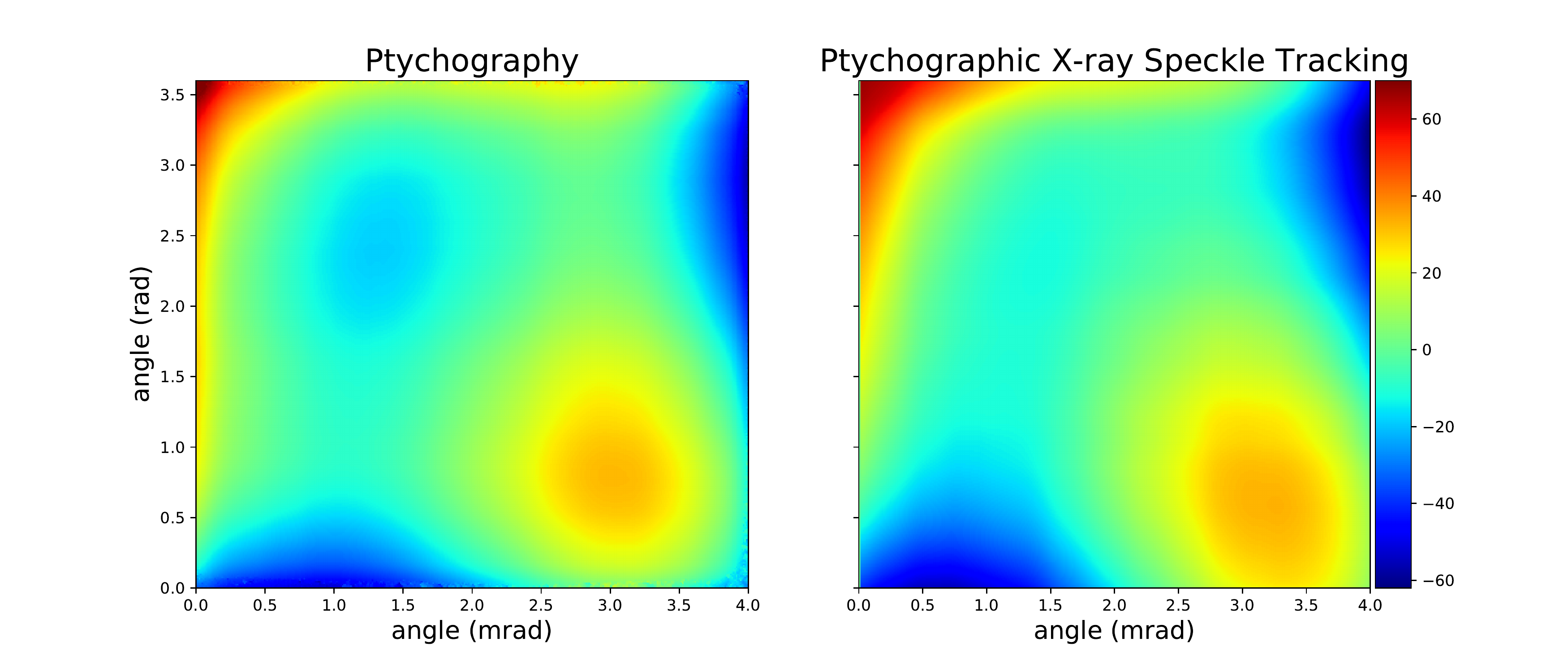}
    \caption{Phase of the recovered wavefronts via ptychography (left) and PXST (right). The phase ptychographic phase profile was unwrapped before display. The colour scale is in radian units.}
\label{fig:comparison}
\end{figure}

In Fig. \ref{fig:comparison} we show the phase of the wavefront recovered via far-field ptychography (left) and PXST (right), from two independent datasets obtained at the European Synchrotron Radiation Facility\footnote{Further analysis using these datasets can be found in \cite{Murray2019a}.}. In both experiments, a Siemens star test sample was scanned in a 2D grid pattern across the wavefront, with a focus-to-sample distance of $0.13\;$m. For the ptychographic dataset, the sample was scanned near the focal plane of the lens ($z_1=1.01\;$mm) and the angular extent of the diffraction extended well beyond that of the diverging illumination, i.e. outside of the holographic region. For the PXST dataset, the sample was placed further from the focus (with $z_1=5.8\;$mm) and the diffraction was predominantly confined to the holographic region of the detector (in a $2\times1.6\;$mrad angular window), consistent with the near-field scattering regime. 

One advantage of PXST over ptychography, is that the phase profile is not ``wrapped'' onto the $[-\pi, \pi)$ domain. This is useful in cases where the intent is for the recovered phases to inform a structural analysis of the lens system, such as the height of a mirror surface or the local period of bi-layers in an MLL. In some cases however, the phases recovered from ptychography can ``un-wrapped''; for smooth phase profiles, continuity of the phases allows one to identify regions bounded by discontinuous $2\pi$ phase jumps. One can then add or subtract $2\pi$ to the phases in these regions as needed until the entire phase profile is smooth. This procedure was applied to unwrap the phases shown in the left panel. 
\onecolumn
\twocolumn
The two phase profiles show qualitative agreement between the ptychographic and PXST algorithms. The root-mean-squared deviation is $\approx 5\;$rad, which is many orders of magnitude worse than theoretically achievable phase sensitivity. Therefore, one or both of the reconstructions suffers from systematic artefacts in the recovered phases. This is a matter for further investigation.

\end{document}